\documentclass[a4paper,11pt]{article}
\usepackage{amsmath, amssymb, amsfonts,indentfirst,graphicx,color,anysize}
\usepackage{hyperref}
\pdfoutput=1
\marginsize{3cm}{3cm}{2cm}{2cm}
\title{
  \includegraphics[width=0.35\textwidth]{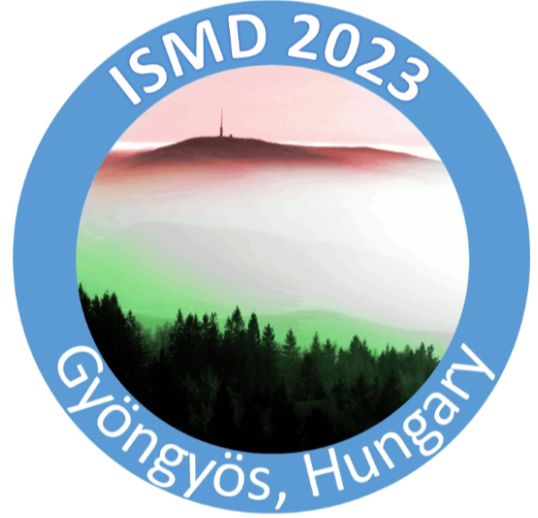}\\[1cm]
  \vskip -0.5cm
   \textbf{Checking the $^8$Be anomaly with a two-arm electron
    positron pair spectrometer }}

\author{{Tran~The~Anh $^{1}$, Tran~Dinh~Trong $^{2}$, Attila J.~Krasznahorkay$^3$,}\\
  {Attila~Krasznahorkay$^3$, József~Moln\'ar$^3$, Zolt\'an~Pintye$^3$, Nguyen~Ai~Viet$^1$,}\\
  {Nguyen~The~Nghia$^{1}$, Do~Thi~Khanh~Linh$^4$, Bui~Thi~Hoa$^1$, Le~Xuan~Chung$^4$ and,}\\ {Nguyen~Tuan~Anh$^5$}\\[1ex]
  $^{1}$ VNU-University of Science, Vietnam National University,\\
  334 Nguyen Trai, Hanoi, Vietnam;\\
  $^{2}$ Institute of Physics, Vietnam Academy of Science and Technology,\\
  18 Hoang Quoc Viet, Hanoi, Vietnam;\\
  $^{3}$ Institute for Nuclear Research  (HUN-REN ATOMKI), \\
  P.O. Box 51, H-4001 Debrecen, Hungary\\
  $^{4}$  Institute for Nuclear Science and Technology,\\
  VINATOM, 179 Nghia Do, Hanoi, Vietnam;\\
$^{5}$  Hanoi Irradiation Center, VINATOM, Cau Dien, Hanoi, Vietnam\\
}

\begin{document}

\maketitle

\begin{abstract}
We have repeated the experiment performed recently by Krasznahorkay et
al., (Phys. Rev. Lett. 116, 042501 (2016)), which may indicate a new
particle called X17 in the literature. In order to get a reliable,
and independent result, we used a different structure of electron-positron
pair spectrometer at the VNU University of Science. The spectrometer has two arms and simpler acceptance/efficiency as a
function of the correlation angle, but the other conditions of the
experiment were very similar to the published ones. We could confirm
the presence of the anomaly measured at E$_p$= 1225 keV, which is above the E$_p$=1040 keV resonance. 
\end{abstract}

\section{Introduction}

An experiment was conducted at the ATOMKI Laboratory (Debrecen,
Hungary) \cite{kr16} in 2016, studying the $^7$Li(p,e$^+$e$^-$)$^8$Be nuclear
reaction. The target nucleus was excited through proton capture, with
the experiment set up to detect e$^+$e$^-$ pairs produced in the Internal Pair
Creation (IPC) during the transition from the excited to the ground
state of $^8$Be. The experimental setup was using a set of multiwire
proportional counters placed in front of $\Delta$E and E detectors, to
determine the e$^+$e$^-$ opening angle,  $\theta (e^+e^-)$. The very thin
$\Delta$E detectors
were made of plastic scintillators and chosen to provide excellent
gamma suppression. While the much thicker E detectors were used to
measure the total energy of the electron and positron. A detailed
description of the experimental setup can be found in \cite{gu16}. The ATOMKI
collaboration observed a deviation in the  $\theta (e^+e^-)$ distribution with
respect to the expected Rose theory distribution \cite{ro49}, at around 140
degrees.

Zhang and Miller \cite{zh21} studied the protophobic
vector boson explanation in $^8$Be, by deriving an isospin relation between the
coupling of photon and X17 to nucleons. They are
expected to have contributions from M1 multipolarity transitions
coming from the resonant proton capture (17.6 and 18.15 MeV $J^\pi =
1^+$ states) as well as from the E1 multipolarity transitions
resulting from the direct proton-capture process.

After 2016 the ATOMKI collaboration repeated the
measurements, while improving the experimental setup in many different
ways~\cite{kr21,kr22}. The anomaly kept appearing in the follow-up experiments,
with no nuclear physics model being able to explain it. This led to
the explanation of a new particle, beyond the Standard Model of
particle physics, created and decaying to e$^+$e$^-$ pairs, being detected
in these experiments. The hypothetical particle is now commonly
referred to as X17, because of the invariant mass calculated from the
e$^+$e$^-$ anomalies.

In addition to the $^7$Li(p,e$^+$e$^-$)$^8$Be nuclear reaction,
since 2019, the same collaboration has also studied the $^3$H(p,e$^+$e$^-$)$^4$He~\cite{kr21}
and $^{11}$B(p,e$^+$e$^-$)$^{12}$C~\cite{kr22} reactions. Different proton energies were used,
leading to anomalies appearing in the  $\theta (e^+e^-)$ distributions at
different angles. All well consistent with the assumption of an $m_X \approx$
17 MeV particle being created with different kinetic energies, leading
to different opening angles between the electron positron pairs.

The
ATOMKI anomaly, being a genuine physics effect, is supported by a number
of arguments.
\begin{itemize}

  \item{}
The anomaly has been observed in $^8$Be with experimental
setups using different geometries, with 5 and 6 arm spectrometers~\cite{kr18}.
\item{}
The anomaly has been observed using  fundamentally different position
sensitive detectors: multi-wire chambers and silicon strip detectors~\cite{kr16, kr18}.
\item{}
The anomaly has been observed in three different nuclei by now ($^9$Be,
$^4$He and $^{12}$C), showing up at e$^+$e$^-$ opening angles consistent with a
single particle~\cite{kr16,kr21,kr22}.
\item{}
The anomaly has been observed at different proton
beam energies at varying e$^+$e$^-$ opening angles, also consistent with a
single particle~\cite{kr16,kr21,kr22}.
\item{}
All observed anomalies have a very high statistical significance such as 6.8$\sigma$~\cite{kr16}, $6.6-8.9\sigma$~\cite{kr21}, and $3-8\sigma$~\cite{kr22}.
\end{itemize}

Despite the consistency of the numerous observations at
ATOMKI, more experimental data is needed to understand the nature of
this anomaly. Many experiments around the world started looking for
such a particle in different channels, or are planning to do so. Many
of these experiments~\cite{al23,ka24,na62,ma23} already put constraints on the coupling of such a
hypothetical particle with ordinary matter. Others are still in an R\&D
phase, soon to contribute to a deeper understanding of this
phenomenon as concluded by the community report
of the Frascati conference \cite{al23}. The report also gives a nice
overview of the possible theoretical interpretation of the observed
anomalies in $^8$Be, $^4$He, and $^{12}$C~\cite{al23}.

At the VNU University of Science (HUS), we have a 5SDH-2 Pelletron accelerator, a 1.7 MV tandem electrostatic type. It can provide a proton beam with energy from 0.35 to 3.4 MeV or 0.7-5.1 MeV for the alpha beam. The accelerator was installed and operated in 2011~\cite{ng11}, and some initial work for the research in nuclear reaction cross-sections has been done~\cite{cu19}. This article aims to look for the 8Be anomaly at the VNU University of Science (HUS) with a two-arm electron-positron spectrometer specifically designed and built for this purpose.

%%%%%%%%%%%%%%%%%%%%%%%%%%%%%%%%%%%%%%%%%%
\section{Experimental methods}

In order to make our experimental results easier to reproduce, in the
first part of this paper, we give a more precise description
of our $e^+e^-$ pair spectrometer with the setup of the detectors
and the electronics connected to them. This will be followed by a
brief description of the data acquisition system and the results of
testing and calibrating the detectors.

\subsection{The e$^+$e$^-$ spectrometer}

\begin{figure}[htb]
  \begin{center}
    \includegraphics[width=10.5 cm]{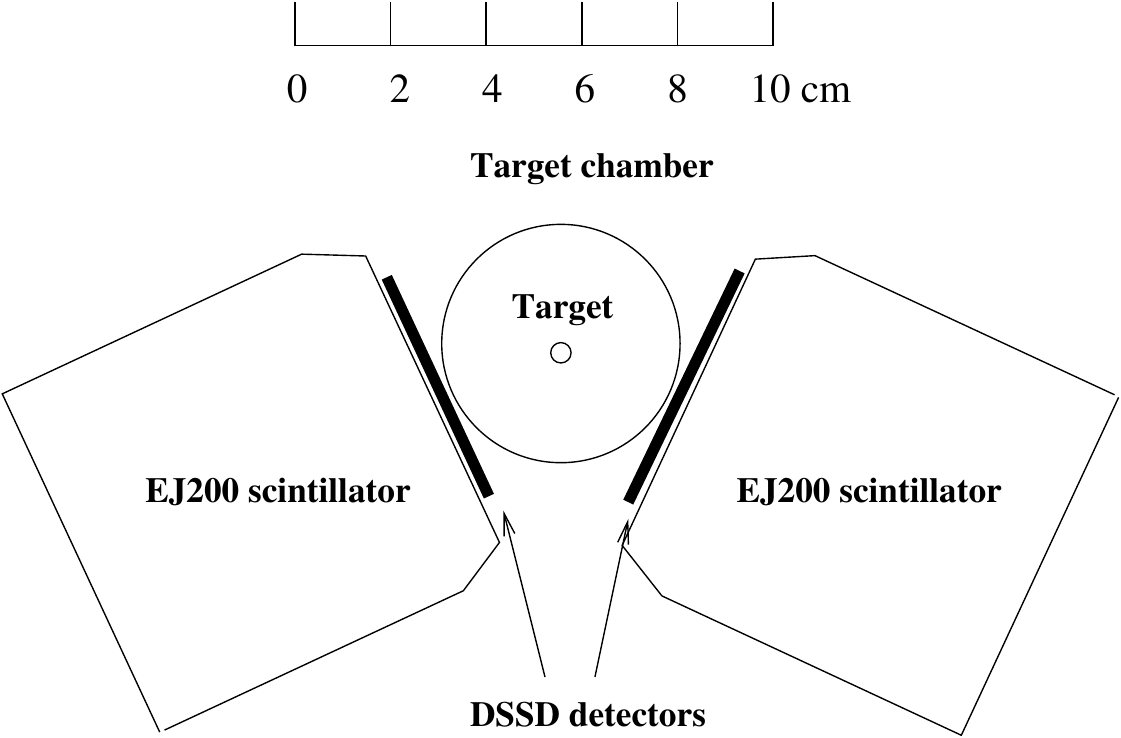}
  \end{center}
%  \vspace{-0.5cm}\begin{figure}[H]
  \caption{Schematic diagram of the e$^+$e$^-$ spectrometer focusing on 140 degree region.
  \label{fig:setup}}
\end{figure}

In the present experiment, two detector telescopes, including Double Silicon Strip Detectors (DSSD)  and  plastic scintillators
were used, placed at an angle
of 140$^\circ$ with respect to each other. The diameter of the carbon
fiber tube of the target chamber has been reduced from 70 mm (used in ATOMKI)
to 48 mm
to allow a closer placement of the telescopes to the target.
This way, we could cover a similar solid angle at 140$^\circ$ like the one
used in the ATOMKI experiment Fig.~\ref{fig:setup}.

\unskip

In this setup, the
efficiency function has only one maximum as a function of the $e^+e^-$
opening angle. This angular dependence can be simulated and calibrated
more robustly than for more complicated configurations.
Another advantage of this setup that we used is that its sensitivity
to background from cosmic radiation is significantly less.

The two detector telescopes were placed at azimuthal angles -20$^\circ$ and -160$^\circ$
with respect to horizontal. This meant that cosmic rays, predominantly arriving vertically,
would have a very low chance of hitting both telescopes at the same time.

For the detection of 1-20 MeV e$^+$ and e$^-$  particles, which are
considered high-energy in nuclear physics, Geant4\cite{al16} calculations were
performed for different detector materials. This included plastic
scintillators, Ge semiconductor detectors, and LaBr$_3$
scintillators. Much better energy resolution could be achieved with
the latter two types of detectors than with a plastic scintillator. An FWHM below 20 keV can be achieved at 17.6 MeV for the Ge detector and 150 keV for the LaBr3 detector. However,
our simulations also showed that in the case of high-density and
high Z detectors, the efficiency of the total energy detection in
the energy response function is greatly reduced compared to the
integral of the response function. The ratio between the numbers of full energy events and total recorded events is smaller than 1.5\% at an electron energy of 18 MeV for a 3x3 inch$^2$ LaBr3 detector.
In materials with higher atomic numbers, the  e$^+$ and e$^-$ particles slow down faster,
and thus, the probability of generating bremsstrahlung radiation is higher.
However, there is a high probability that these radiations will escape the detector.
This is why the probability of detecting  e$^+$ and e$^-$ particles at full energy is reduced in
these high Z materials.
Therefore, we used
special plastic
scintillators for this task. The dimensions of the EJ200 plastic
scintillators were chosen (82 × 82 × 80 mm$^3$ each) in such 
a way that these high-energy
particles would be completely stopped in them.

To collect the light generated by the e$^+$e$^-$
particles slowing down in the detector from all corners with the same
efficiency, the SCIONIX company provided the detector with specially shaped
light guides. The surfaces of both the
detectors and the light guides were diamond polished. The collected light
was converted to an electronic signal by Hamamatsu photomultiplier tube 
(PMT) type R594 assemblies.

\begin{figure}[htb]
  \begin{center}
    \includegraphics[width=7.5 cm]{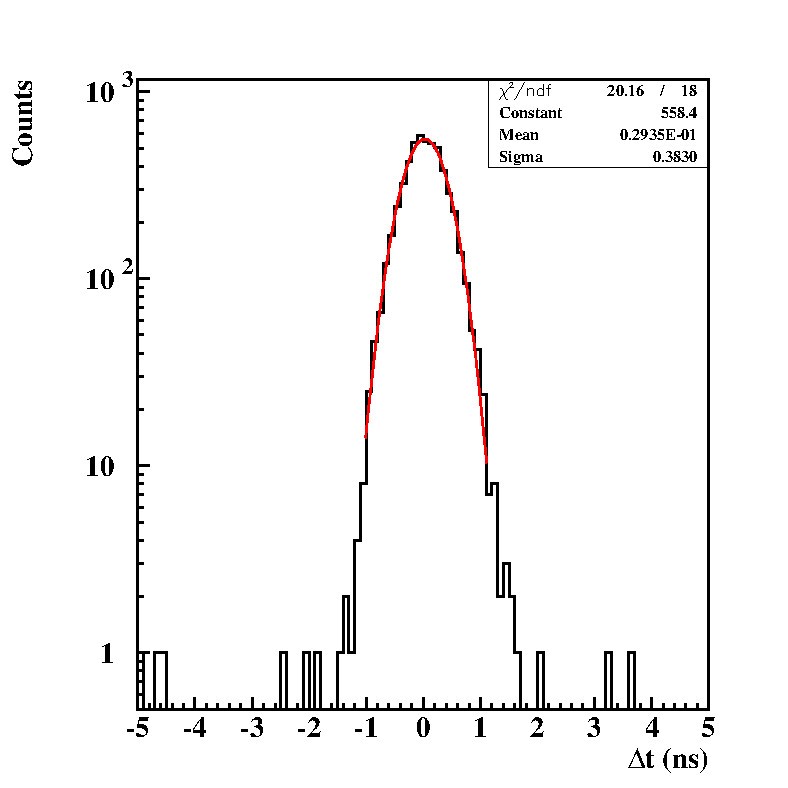}
  \end{center}
%  \vspace{-0.5cm}\begin{figure}[H]
  \caption{Time difference distribution between the two plastic scintillators using two cascade-gamma rays emitted from $^{60}$Co source.
  \label{fig:t-plastic}}
\end{figure}

The time resolution of the detector turned out to be adequate despite
the large size of the detector and multiple light reflections
inside. It was measured with a $^{60}$Co source between 2 detectors, and
found to be less than 1 ns, as shown in Fig.~\ref{fig:t-plastic}.

\unskip

In the absence of a high-energy electron source, the energy resolution
of the detector could only be determined with the help of e$^+$e$^-$ pairs
from internal pair creation. We will come back to this in section 2.4 in
connection with the energy calibration of the detectors.

Double-sided Silicon Strip Detectors (DSSD), W1(DS)-500, were used to measure the
energy loss of the $e^+e^-$ particles and their directions.
The DSSDs were purchased from Micron Semiconductor. They consist of 16
sensitive strips on the junction side and 16 orthogonal strips on the ohmic
side. Their element pitch is 3.01 mm for a total coverage of 49.5 x
49.5 mm$^2$. They are mounted on a printed circuit board (PCB), with 34 pins on one edge for
their readout. These are connected via 34-conductor
flat cables to the pre-amp boxes.

Mesytec MUX32 type 32-channel preamplifiers,
linear amplifiers, timing filter amplifiers, timing discriminators and
multiplexers were used. The full width of the 5$^{th}$-order shaped energy signals is 1.5 $\mu$s, and the full rate capability of the MUX32 unit is 800 kHz. The energy resolution of the channels is 5.5 keV Si + 0.064 keV/pF. The timing filter amplifier signals are shaped with 20 ns integration time and 100 ns differentiation time, followed by leading edge discriminators. It is a very fast multiplexed preamplifier, shaper, and discriminator combination with very good energy and timing resolutions. 
The MUX-32 consists of two MUX-16s; each MUX-16 manages 16 inputs, up to two simultaneous responding channels are identified, and two amplitudes plus the two corresponding amplitude-coded addresses (position signals) are sent to the outputs. Therefore, it can manage the double-hit events with full energy and position information in the x and y directions. These modules are especially well-suited for DSSD detectors. 
 
By properly shielding and grounding the detectors, using 10 $\mu$m thick Al
foils mounted on PCBs both at the front and the back side of the
detectors and shielding the flat cables and connectors, we managed to
reduce their electronic noise and could lower the levels of the discriminators
below 50 keV.  For simplicity, the detectors are operated
in air without dedicated cooling. The first test and calibration
of the detectors
were performed with a mixed $\alpha$-source containing $^{239}$Pu, $^{244}$Cm,
and $^{241}$Am. The experimental configuration was used in Geant4 simulation code\cite{al16}. All materials used to construct the experiment have been included in the simulation geometry. Therefore, the energy loss in the Al foil and in the air (5 mm) were taken into account.

\subsection{Data acquisition system}

The data acquisition system uses Versa Module Eurocard (VME), Analog-to-Digital Converter (ADC), Time-to-Digital Converter (TDC), and Charge-to-Digital Converter (QDC) units, which were read out with the help of a commodity desktop PC.
The software used for configuring the VME devices and recording the data
collected can be found in \cite{ati}. 

\begin{figure}[htb]
  \begin{center}
    \includegraphics[scale=0.25]{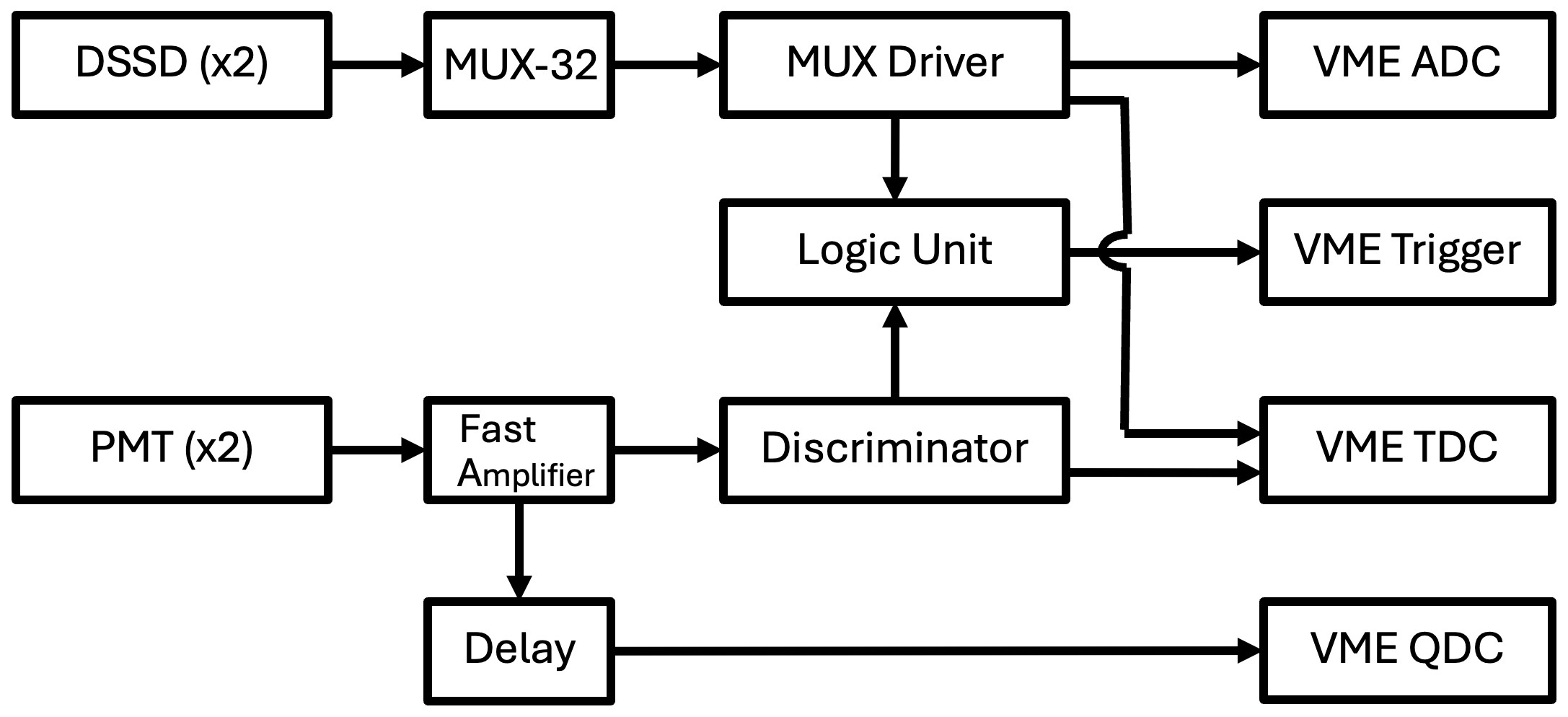}
  \end{center}
%  \vspace{-0.5cm}
  \caption{Electronic block diagram of the e$^+$e$^-$ spectrometer.}
  \label{fig:elek}
\end{figure}

Constant Fraction Discriminators (CFDs) and TDC units were
used to determine the arrival time of the
signals coming from the plastic scintillators, and QDCs were used to digitize their energy signals. The block diagram of the electronics connected to
the detectors is shown in Fig.~\ref{fig:elek}.
The signals received from the MUX32
multiplexer were digitized with the help of ADC and TDC units.

\unskip

\subsection{Calibration of the DSSD detectors}

\begin{figure}[htb]
  \begin{center}
    \includegraphics[scale=0.3]{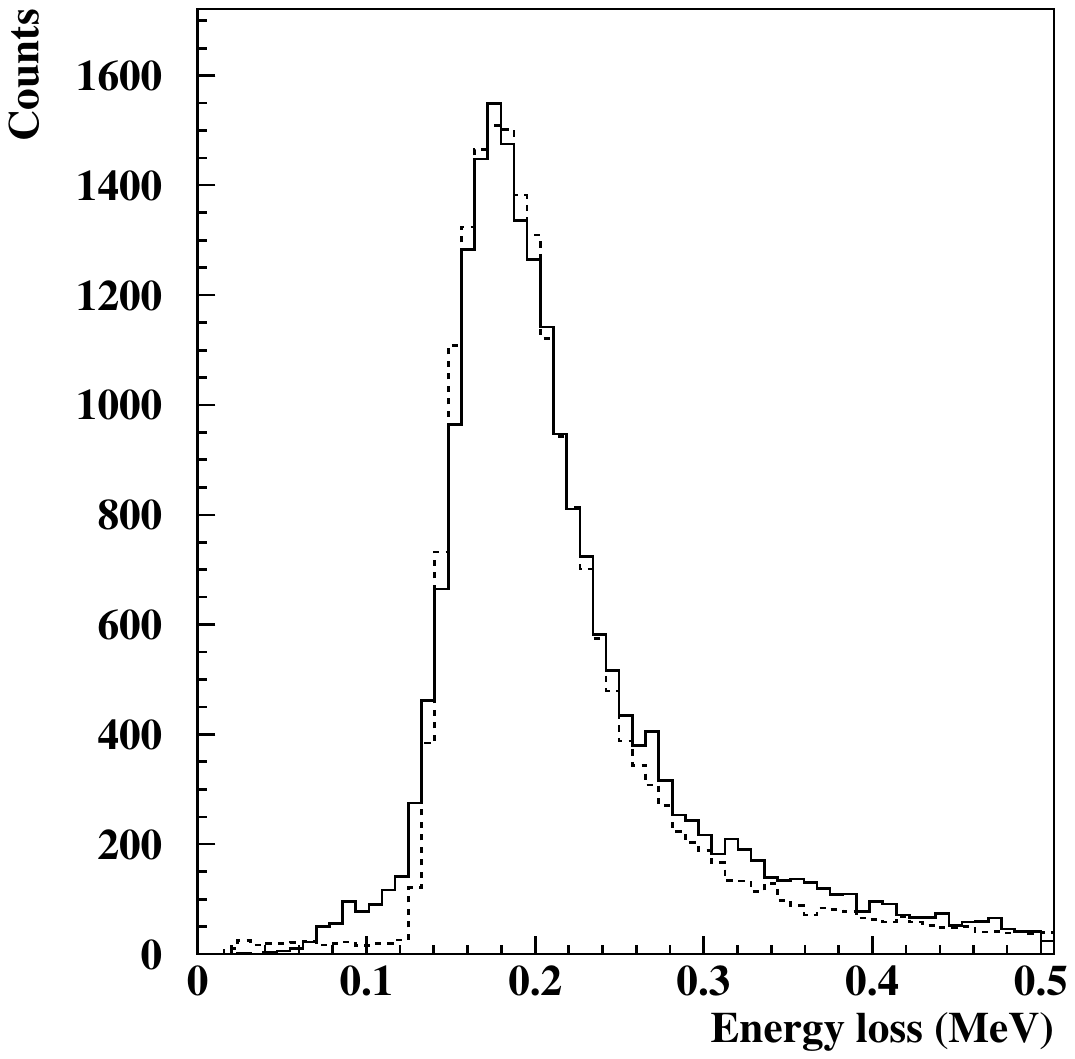}
  \end{center}
%  \vspace{-0.5cm}
  \caption{Typical measured (black histogram)  and simulated (dashed line histogram) energy loss distributions of
electrons and positrons passing through one of the DSSD detectors.}
  \label{fig:de}
\end{figure}
\unskip

The energy measured by the DSSD was calibrated using information from the e$^+$e$^-$ particles, which were produced by the $^{7}$Li($p$,$\gamma$)$^{8}$Be
  nuclear reaction.
The energy of the bombarding protons was set to the
 E$_p$=441 keV resonance. 
 When creating the spectrum, we required real coincidence between the DSSD detector and the plastic scintillator located behind it by requiring a time difference smaller than 400 ns.
 %When creating the spectrum, we required real coincidence between the DSSD detector and the plastic scintillator located behind it by requiring the time different between two plastics is smaller than 40 ns. 
 The two plastics measured the total energy of the  e$^+$e$^-$ particles.
We also required this energy to be in the 6-20 MeV
range. The experimental configuration was used in Geant4 simulation code\cite{al16}. 
Fig.~\ref{fig:de} shows the energy loss distribution of
electrons and positrons passing through one of the DSSD detectors.
The histograms show reasonably
good agreement between the experimental and simulated distributions, however
 we obtained some differences around 100 keV and 350 keV.
The former can be
caused by some electronic noise, while the
latter is caused by the events when both the  e$^+$ and e$^-$ created during the internal pair production passed through the same DSSD detector.
The detection efficiency
with a CFD threshold of 50 keV was greater than 97\% for electrons and
positrons. The 120 keV lower limit on the electron / positron energy applied in the event selection.

\begin{figure}[htb]
  \begin{center}
    \includegraphics[scale=0.3]{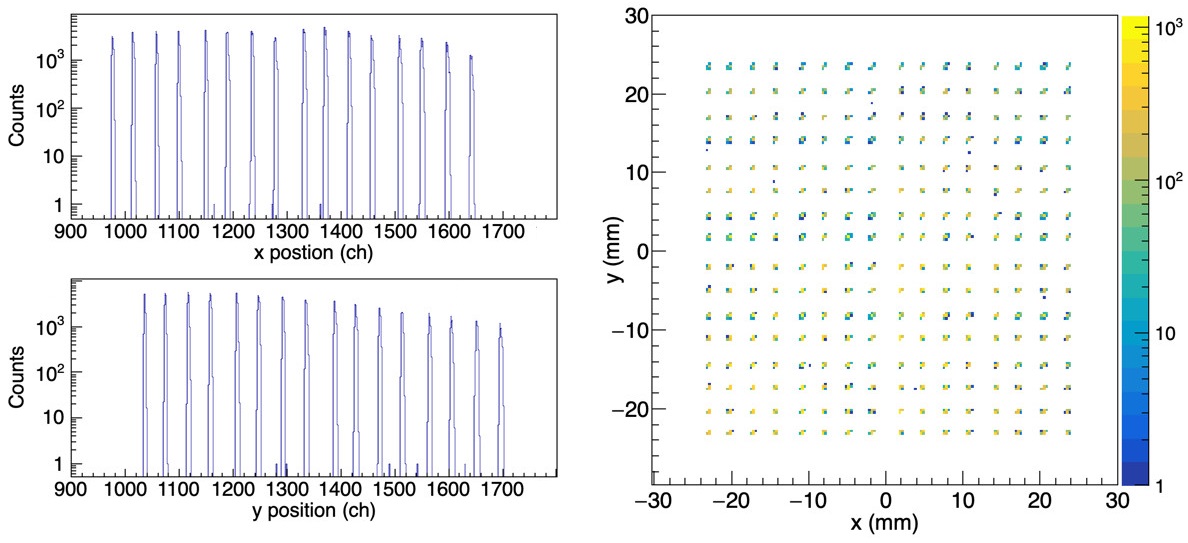}
  \end{center}
%  \vspace{-0.5cm}
  \caption{Left side: the x and y distribution of the position signals obtained
    from one of the DSSD detectors. Right side: 2D distribution of the
    position signals.}
  \label{fig:dssdxy}
\end{figure}

The histograms in Fig.~\ref{fig:dssdxy}  show the (non calibrated) distribution of the x and y coordinates of the
impact points of the e$^+$/e$^-$ particles that hit the DSSD detector. The
peaks of the fence spectrum correspond to the coordinates of
particles passing through the individual silicon strips. As shown, the assignment
of the recorded data to the x and y strips is very clear. The width of DSSD was used to convert the recorded data to position in cm scale with origin at the center (two dimension plot in Fig.~\ref{fig:dssdxy}). Of the 70,000 events
recorded for these distributions, only around 3\% of events had to be excluded due to missing information in either the x or y direction.

\unskip

\subsection{Calibration of the scintillation detectors}

\begin{figure}[ht!]
  \begin{center}
    \includegraphics[scale=0.7]{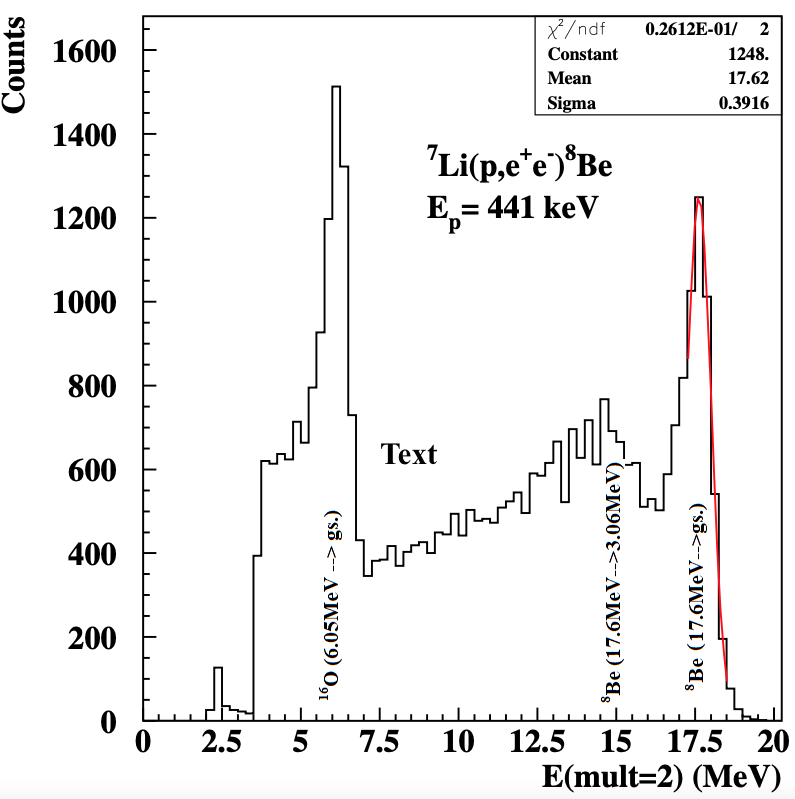}
  \end{center}
  \caption{%The total energy is deposited in a plastic scintillation of the $e^+e^-$pairs, which are selected by double-hit events in the corresponding DSSD. The peaks are from the transition indicated in the figure; the red line is a fitting line of a Gaussian with its center and sigma shown in the legend.
               The total energy deposited in the plastic scintillator by $e^+e^-$ pairs, which are selected as double-hit events in the corresponding DSSD. The peaks are from the transitions indicated in the figure; the red line is a Gaussian fit with its center and sigma shown in the legend. }
  \label{fig:sume}
\end{figure}

The energy distribution of the electrons and positrons in internal
pair creation individually is contiguous.
However, if the electron and positron lose their
energy in the same detector, the energy distribution of such events will
show peaks at the transition energies - $2m_ec^2$. The scintillator
detectors were calibrated using such events.
If the detectors are close enough to the
target, there is a high probability of the above internal energy summing.
The energy spectrum measured by the scintillators for events selected by gating on double-hit in the DSSD detector, can be found in Fig.~\ref{fig:sume}.

\unskip
%%%%%%%%%%%%%%%%%%%%%%%%%%%%%%%%%%%%%%%%%
\section{Experimental results}

The experiments were performed in Hanoi (Vietnam) at the 1.7 MV
Tandem accelerator of HUS, with different proton beam energies between
0.4$\leq E_p\leq 1.3$ MeV. The typical beam currents for these experiments were from 1 to 1.5 $\mu$A.

LiF targets with thicknesses of $\approx$30 $\mu$g/cm$^2$ evaporated onto 10 $\mu$m thick Al foils, as well as Li$_2$O targets with thicknesses of $\approx$0.3 mg/cm$^2$, were used on 1  $\mu$m thick Ni foils in order to maximize the yield of the $e^+e^-$pairs. The LiF is a more stable target, and it is easy to evaporate. That was the reason we used it at low bombarding energy (E$_p$=441 keV, E$_x$=17.6 MeV) to calibrate the spectrometer. However, if we increase the beam energy from 441 keV (E$_x$= 17.6 MeV) to 1.04 MeV (E$_x$=18.15 MeV), the cross-section of the $^{19}$F(p,$\alpha$)$^{16}$O reaction increases very fast and the created e$^+$e$^-$pairs coming from the decay of the 6.05 MeV E$_0$ transition would overload our electronics and data acquisition, and observing e$^+$e$^-$pairs from the 18.15 MeV transition would not be feasible.

$\gamma$ radiations were detected by a 3''x3'' NaI(Tl) detector
monitoring also any potential target losses. The detector was placed
at a distance of 25 cm from the target at an angle of 90 degrees to
the beam direction.

A single energy spectrum measured by the scintillators and gated by
``multiplicity=2'' events in the DSSD detector, which means that both
the electron and positron coming from the internal pair creation are
detected in the same telescope,  is shown in Fig.~\ref{fig:sume} for telescope~1.
Figure~\ref{fig:sume} clearly shows the transitions from the
decay of the 17.6 MeV resonance state to the ground and first excited states
in $^{8}$Be. The
cosmic ray background was already subtracted from that. The energy resolution of the plastic scintillator at 17.6 MeV was extremely good (5.2\% FWHM), proving the very good light collection from the whole detector. The ‘’background” below the peak comes mostly from the wide (1.5 MeV) transition going to the first excited state of $^{8}$Be and from the tail of the 17.6 MeV transition. The energy resolution of the 6.05 MeV ($^{16}$O) peak is also good.

\begin{figure}[ht!]
  \begin{center}
    \includegraphics[scale=0.3]{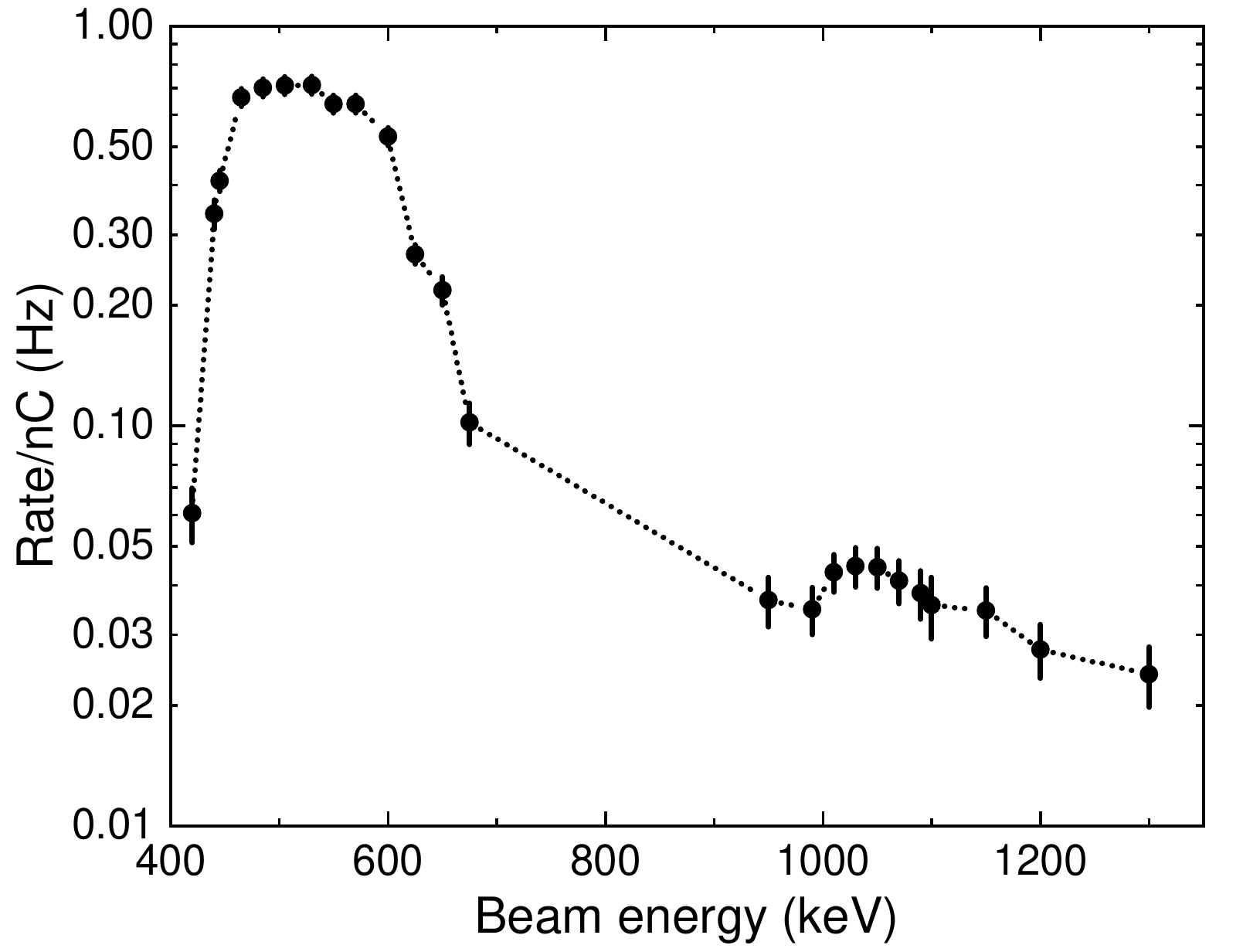}
  \end{center}
  \caption{The excitation function of $^8Be$ was obtained by scanning proton energy; the resonances of 17.6 and 18.15 MeV were seen as proton beams around 500 and 1040 keV, respectively. The measured data are shown by black dots with error bars and the dotted line connected the points is just drawn to guide the eye.}
  \label{fig:thickness}
\end{figure}
\unskip

In order to check the effective thickness of the Li$_2$O targets, we measured
the excitation function of $^8$Be via $^7$Li(p,$\gamma$)$^8$Be
reaction by scanning the proton beam energies from 441 keV to 1300
keV. The events with multiplicity-2 in DSSD and measured energy in 
the plastic scintillator larger than  10 MeV were counted, and their rate was plotted in
Fig.~\ref{fig:thickness}. Two resonance peaks were observed
at the proton beam energies of around 441 and 1040 keV \cite{ti04,za95}. The
width of the resonance shows the effect of the target thickness. For the 441 keV resonance, it was found to be approximate 150 keV, which means about 0.44 mg/cm$^2$ target thickness, since the energy loss of the protons is 340 keV/mg/cm$^2$.

\begin{figure}[ht!]
  \begin{center}
    \includegraphics[scale=1]{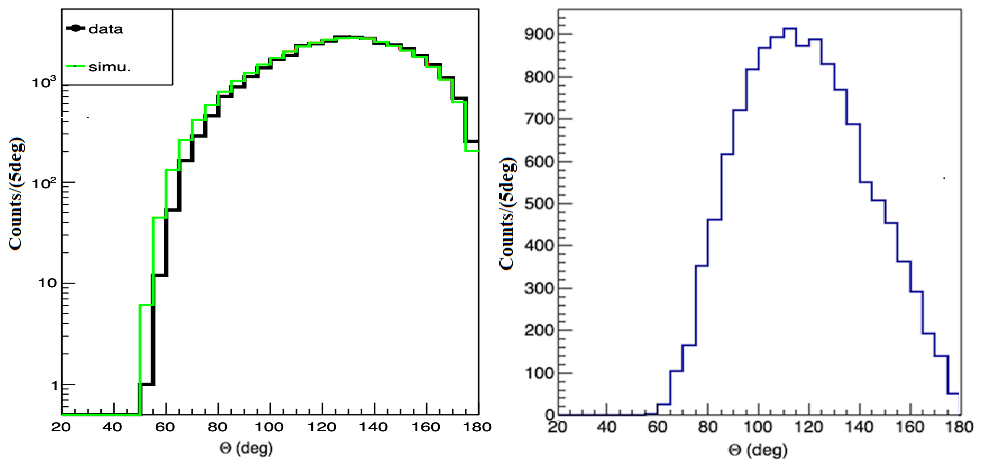}
  \end{center}
  \caption{The acceptance curve of two-arm telescope system (left) and angular correlation of e$^+$e$^-$ pairs obtained from the 17.6 MeV transition using LiF target (right).}
  \label{fig:acceptance}
\end{figure}
\unskip
The efficiency (acceptance) as a function of the correlation angle in
comparison to isotropic emission was determined from the same data set
by using uncorrelated e$^+$e$^-$ pairs formed of separate, single
events\cite{gu16}. To do this,
uncorrelated e$^+$e$^-$ pairs have been recorded during the experiment.
The analysis would select event pairs with uncorrelated electrons/positrons
hitting different telescopes in the events.
The opening angle distribution of electron/positron pairs from such
events is shown by the black line in Fig.~\ref{fig:acceptance} (left). The
green line is a simulation curve produced using Geant4. They show quite a
good agreement between the estimated experimental and simulation efficiencies.

Coincidence events, with both arms of the spectrometer detecting e$^+$/e$^-$particles, were also recorded. The opening angle distribution of e$^+$e$^-$ pairs from such events is shown in Fig.~\ref{fig:acceptance} (right). The cosmic background data had been collected and analyzed similar to the experiment data and subtracted. The total time collection of both data had been normalized.

\begin{figure}[ht!]
  \begin{center}
    \includegraphics[scale=0.74]{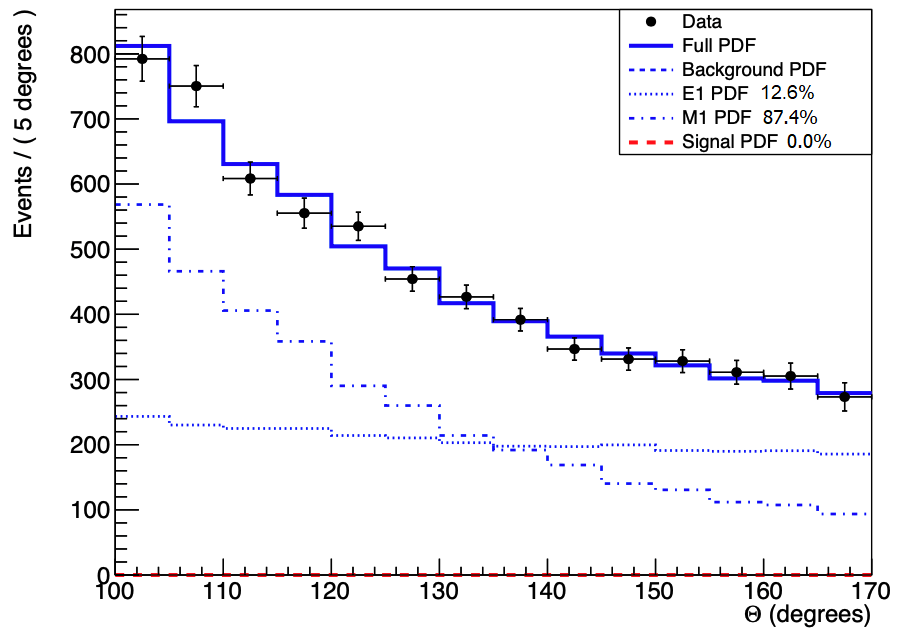}
  \end{center}
  \caption{Angular correlation of e$^+$ e$^-$  pairs when the LiF target was bombarded by a proton beam at 441 keV. The background line is the summing up of E1 and M1 lines, the numbers in the legend give the contributions of each component. Both the experimental data and simulations are corrected by the detection efficiency. }
  \label{fig:e441}
\end{figure}
\unskip

In the first experiment, we used a proton beam energy of 411~keV to bombard
the LiF target. Through this, the $^8$Be nucleus would be created in the
17.6~MeV excited state.
Figure~\ref{fig:e441} shows the angular correlations of e$^+$e$^-$
pairs originating from the transition of this 17.6~MeV excited state
of $^8$Be to its ground state. The Monte Carlo detector simulations of the
experiment were done using Geant4 and are shown as histograms in
Fig.~\ref{fig:e441} for M1 (dash-dotted line) and E1 (dotted line) multipolarity transitions.
The simulation included the geometries of the target chamber,
target backing, and detector arm assemblies. The interaction of generated
electrons, positrons, and gamma rays was then simulated with the
experimental setup. Internal Pair Creation (IPC) events, generated from
both the possible E1 and M1 transitions, were simulated this way.
The combination of the E1+M1 distributions shows a good agreement with
the experimental data when the contributing fractions are fitted. The dominant M1 part (87.4\%) is clearly understood since it is a 1$^+$ --> 0$^+$ transition. The 12.8\% E1 mixing can also be understood
since the energy loss in the target was about 150 keV, which is about 14 times larger than the width of the resonance ($\Gamma$=10.7 keV)~\cite{ti04},  and we integrated a reasonable amount from the proton direct capture part of the excitation function~\cite{za95} as well, which has a multipolarity of E1.

\begin{figure}[ht!]
  \begin{center}
    \includegraphics[scale=0.74]{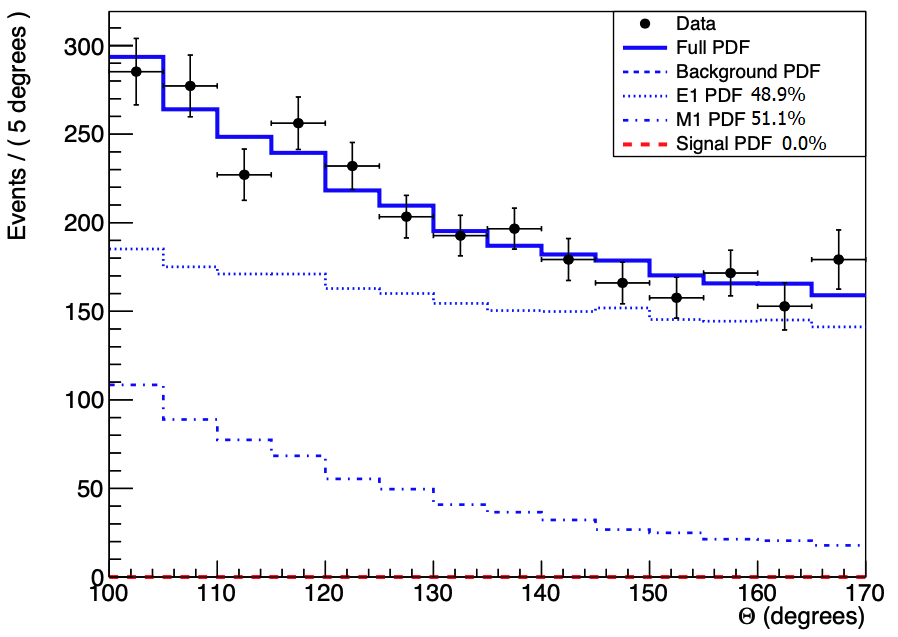}
  \end{center}
  \caption{Similar to Fig.~\ref{fig:e441} with 800 keV proton beam bombarding on LiF target.}
  \label{fig:e800}
\end{figure}
\unskip
In the second experiment, we changed the proton beam energy to 800 keV
at the off-resonance energies.  With the same method to build the
total simulation curve and show the experimental data in the
Fig.~\ref{fig:e800}, we can see the simulated curve go through the
middle of data points. There is no systematic deviation of the
experimental points from the IPC simulation curve. As can be seen in the insert of Fig 10, the background could be described well with 48.9\% E1 and 51.1\% M1 components. The E1 component comes from the direct proton capture, while the 51.1\% M1 component comes from the tails of the E$_p$=441 keV and 1040 keV resonances. We did not observe any contribution from the X17 decay like N.J. Sas, et al.~\cite{sa22}  observed before. Since during this experiment the target was burned out (punctured) many times, the effective energy of the protons was changing and may have washed out the anomaly caused by the X17 to e$^+$e$^-$ decay.

\begin{figure}[ht!]
  \begin{center}
    \includegraphics[scale=1]{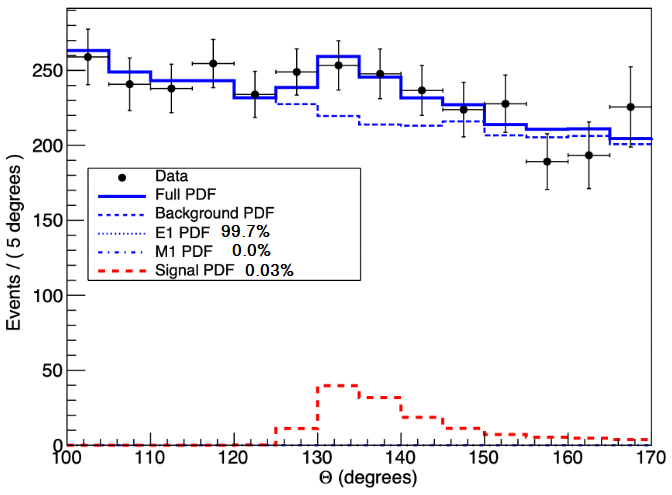}
  \end{center}
  \caption{Angular correlation of e$^+$e$^-$  pairs when the Li$_2$O target was bombarded by a proton beam at E$_p$= 1225 keV. The background line is the summing up of E1 and M1 lines, the contribution of M1 transition is negligible,                                     so the background line is overlapped with E1 line.}
  \label{fig:e1040}
\end{figure}
\unskip
Finally, we changed the proton beam energy to 1225 keV, above to the 1040 keV resonance to check the off resonance region. The combined IPC
simulation curve and experimental data for this transition is shown in
Fig.~\ref{fig:e1040}. It clearly shows a deviation in the e$^+$e$^-$
opening angle distribution between the data and the simulation around
$135^{\circ}$.
This deviation is in agreement with the result
published by the ATOMKI collaboration in~\cite{kr16}. 
By assuming that the deviation is coming from the creation and immediate decay of an intermediate particle to an e$^+$e$^-$ pair~\cite{ro49}, we can calculate a mass of
$m_Xc^2=16.66 \pm 0.47$~(stat.)~MeV for this particle with a confidence
above 4$\sigma$. As can be seen in the insert of Fig~\ref{fig:e1040}, we could describe the background with pure E1 distribution, which show that we are indeed in the off resonance region.

%The difference in the angle of the anomaly obtained in this study of around $135^{\circ}$, and the published one of $140^{\circ}$\cite{kr16}, could be explained by a slight difference in the proton beam spot position on the target. Furthermore, the angle of the anomaly is difficult to define and is only an approximate value. The effect of the uncertainty of the beam position was included in the systematic error. 
The systematic uncertainty on the calculated particle mass from the beam spot’s position was estimated using a series of simulations using different beam spot positions. This resulted in a $\Delta
m_Xc^2$(systematic)= $\pm 0.35$~MeV uncertainty.
Additional potential systematic uncertainty caused by e$^+$e$^-$ pairs induced by external pairing in DSSD detectors by gamma radiation is being evaluated in simulation.

Based on the best-fit results shown in Fig.~\ref{fig:e1040}, it can be concluded that X17-boson particles were created simultaneously with the IPC  due to the E1 transition in this experiment. The branching ratio of the e$^+$e$^-$decay of such boson to IPC and $\gamma$ decay of the 18.15 MeV level is found to be 2.8x10$^{-3}$ and 1.1x10$^{-5}$, respectively. 
It seems that the X17 particle is created in the E1 transition and not in the M1 one. In Ref.~\cite{kr16}, they obtained a branching ratio of 5.8x10$^{-6}$, which is about half of the value we obtained here. They did the experiment on the 1040 keV resonance, in this way the M1 contribution of the resonance may not produced any X17 particle.

\section{Summary}

We successfully built a two-arm e$^+$e$^-$ spectrometer in Hanoi. The
spectrometer was tested and calibrated using the 17.6~MeV M1
transition excited in the $^7$Li(p,e$^+$e$^-$)$^8$Be reaction. We have
got a nice agreement between the experimentally determined acceptance of
the spectrometer with the one coming from our simulation.
The angular correlation of the e$^+$e$^-$ pairs measured for the 17.6 MeV transition (E$_p$=441keV) agrees well with the simulated one dominated by  the M1 transition, and no anomaly was observed for this. However, a significant anomaly 
($>4\sigma$) was observed for  E$_p$=1225 keV, which is the off resonance region above the E$_p$=1040 keV resonance, at around  135$^{\circ}$, in agreement with the ATOMKI results published
in 2016\cite{kr16}. The mass of the hypothetical particle from this result
was obtained to be $16.66 \pm
0.47$(statistical) $\pm 0.35$(systematic)~MeV. In the future,
we are planning to upgrade the
spectrometer to get a wider angular acceptance.

\end{document}